\documentclass[final,1p]{elsarticle}

\usepackage{graphicx}
\usepackage{amsmath,amssymb,amsfonts}
\usepackage{booktabs}
\usepackage{siunitx}
\usepackage{xcolor}
\usepackage{url}
\usepackage{tikz}
\usepackage{microtype}
\usetikzlibrary{arrows.meta,positioning,fit,backgrounds,calc}
\usepackage{listings}
\graphicspath{{figures/}}
\providecommand{\botrule}{\bottomrule}  
\newcommand{\optixreorder}{\texttt{optixReorder}}
\newcommand{\warpeff}{warp execution efficiency}
\newcommand{\Warpeff}{Warp execution efficiency}

\journal{Computer Physics Communications}

\begin{document}

\begin{frontmatter}

\title{Mitigating warp divergence in GPU optical photon Monte Carlo: an order of magnitude speedup with Shader Execution Reordering}

\author{Gabor Galgoczi}
\ead{gaborgalgoczi@gmail.com}  
\affiliation{organization={Brookhaven National Laboratory},
             city={Upton}, state={NY}, postcode={11973}, country={USA}}

\begin{abstract}
In high energy, nuclear, and medical physics, optical photon transport Monte Carlo is a frequent bottleneck in detector simulation. GPU ray tracing accelerates photon propagation, but one thread per photon megakernels suffer from SIMT execution divergence when photon lifetimes vary widely. Short-lived photons leave inactive lanes while a few long-lived photons delay warp completion. We evaluate NVIDIA Shader Execution Reordering (SER) in an OptiX based optical photon transport kernel and use it to regroup surviving photons during propagation. In a \SI{14.7}{\kilo\tonne} liquid argon time projection chamber benchmark with ${\sim}61$ million photons from a \SI{2.5}{\giga\electronvolt} electromagnetic shower, SER capable execution reduces propagation kernel cycles by \num{16.55(23)}$\times$ and end-to-end optical simulation time, including transfers and initialization, by \num{14.72(30)}$\times$ with bit identical hit output. We decouple the speed-up into a \num{3.2}$\times$ contribution from the launch configuration changes introduced when SER is enabled and a further \num{5.1}$\times$ from executing the reorder. Profiling identifies active lane recovery as the dominant mechanism, while branch efficiency and memory bandwidth change little. A volume coherent SER hint enables an analytic navigation fast path in bulk argon, adding a further \num{1.13}$\times$ wall-time speedup. Utilizing SER resulted in a ${\sim}90\%$ reduction in GPU energy per simulated event.
\end{abstract}


\end{frontmatter}

\section{Introduction}\label{sec:intro}

GPU Ray Tracing (RT) has become a broadly adopted accelerator for scientific Monte Carlo simulation. In high energy physics, optical photon transport has been offloaded to GPU ray tracing for nearly a decade: Opticks, built on NVIDIA OptiX, was first demonstrated in 2017~\cite{blyth2017} and has continued to be developed~\cite{blyth_opticks,blyth2025}; it has been used by JUNO~\cite{blyth2020juno}, LZ~\cite{creaner2021lz}, NEXT~\cite{next2025opticks}, and the Electron Ion Collider~\cite{galgoczi2025eic}. The Chroma simulator~\cite{seibert2011chroma} is used by DARWIN~\cite{althueser2022darwin}. GPUs likewise accelerate Monte Carlo in medical physics, specifically photon propagation in tissue~\cite{fang2009mcx}, mesh based transport on OptiX RT cores~\cite{rtmmc}, coupled photon, electron, positron, and neutron transport~\cite{lee2024rt2}, and reactor neutron transport~\cite{salmon2019openmc}. A recent survey catalogues the rapid growth of ray tracing cores for general purpose computing~\cite{meneses2026rtcores}. Reliable and fast simulation underpins detector design and event reconstruction, so accelerating it is of broad importance.

In liquid argon neutrino detectors, prompt scintillation light provides event timing and triggering information complementary to the ionisation charge readout. Liquid Argon Time Projection Chambers (LArTPCs) read out scintillation and Cherenkov light. Their design and reconstruction workflows require repeated, geometry dependent optical photon Monte Carlo: propagating $10^{7}$--$10^{9}$ \SI{128}{\nano\meter} photons per event through wavelength shifting (WLS) layers and onto photosensors. On CPUs this optical stage dominates the cost of a full detector simulation. GPU ray tracing engines built on NVIDIA OptiX, in particular Opticks~\cite{blyth_opticks}, pair GPU ray tracing of the geometry with Monte Carlo sampling of the optical physics, reaching throughputs of order $10^{8}$ photons per second, making explicit optical Monte Carlo practical~\cite{companion}.

Optical photon transport is a particularly suitable workload for this kind of acceleration. Unlike full electromagnetic shower transport on the GPU, for example AdePT~\cite{adept} and Celeritas~\cite{celeritas} where each step can invoke a large and heterogeneous set of physics processes and produce secondary particles, optical photon transport involves only a small, bounded set of per step processes: Rayleigh scattering, bulk absorption, boundary reflection and refraction, wavelength shifting reemission, and detection, with no secondary particle cascade. This keeps the transport a compact megakernel and makes the dominant source of divergence the spread in photon lifetimes: the number of bounces before a photon terminates. That makes it an ideal workload for hardware thread reordering, which directly mitigates lifetime induced divergence.

The GPU transport kernel is a \emph{megakernel}: one kernel carries each photon through its full history, repeatedly tracing the geometry and applying scattering, absorption, boundary, detection, or wavelength shifting physics until the photon terminates. On SIMT GPUs the 32 threads of a warp issue instructions together. If photons assigned to the same warp have very different lifetimes, for example, one is absorbed promptly, another scatters many times in argon, and a third becomes trapped in a wavelength shifting layer then the warp remains resident until the slowest lanes complete, while completed lanes are inactive. This lifetime induced underutilisation is a form of execution divergence and is a well known limiter of GPU ray traversal~\cite{aila2009,laine2013}. Scientific optical Monte Carlo is a particularly severe case because the photon path length distribution can span several orders of magnitude.

Shader Execution Reordering (SER), introduced with NVIDIA Ada GPUs and exposed in OptiX as the \optixreorder{} call~\cite{ser_whitepaper}, allows the OptiX runtime to regroup in flight threads before they continue execution. In rendering workloads this mechanism is used to improve coherence among rays that would otherwise follow different material or traversal paths. To our knowledge SER has so far been applied almost exclusively in real time computer graphics; we are not aware of any prior use in a scientific ray tracing workload. We therefore investigate whether the same mechanism is effective in a scientific Monte Carlo workload whose divergence is dominated not by visual material complexity but by stochastic photon lifetimes.


This paper makes three contributions. First, it reports a comparison of the kernel with and without SER (same seeds) for a \SI{14.7}{\kilo\tonne} LArTPC optical benchmark and identifies the dominant improvement as recovery of active lanes, not a change in branch or memory coalescing behaviour (Section~\ref{sec:mechanism}). Second, it introduces a hybrid ray tracing -- analytical approach for photons deep inside the liquid argon volume with a region coherent reorder hint that preserves warp coherence across an analytic fast path (Section~\ref{sec:r2coh}). Third, it improves the energy usage of the GPU by ${\sim}90\%$ (Section~\ref{sec:discussion}). The underlying optical simulation was validated against Geant4~\cite{companion}.  

\section{Methods}\label{sec:methods}

\subsection{GPU optical photon transport}\label{sec:methods_gpu}

\begin{figure}[t]
\centering
\resizebox{\linewidth}{!}{%
\begin{tikzpicture}[font=\footnotesize,>={Stealth[]},
  proc/.style={draw,rounded corners,align=center,minimum height=9mm,inner sep=4pt,fill=black!3},
  io/.style={draw,align=center,minimum height=9mm,inner sep=4pt}]
  \node[io] (host) {Geant4 (CPU)\\EM shower\\$\rightarrow$ optical photon distributions};
  \node[proc,right=20mm of host] (reord) {\optixreorder\\(every $N$)};
  \node[proc,right=12mm of reord] (trace) {trace BVH\\(${\sim}70\%$)};
  \node[proc,right=12mm of trace] (phys) {optical physics\\scatter / absorb /\\WLS / boundary / detect};
  \node[io,right=18mm of phys] (hits) {hits};
  \draw[->] (host) -- node[above,font=\scriptsize]{\texttt{optixLaunch}} node[below,font=\scriptsize]{1 thread/$\gamma$} (reord);
  \draw[->] (reord) -- (trace);
  \draw[->] (trace) -- (phys);
  \draw[->] (phys) -- node[above,font=\scriptsize]{done} (hits);
  \coordinate (lb) at ($(phys.south)+(0,-0.9)$);
  \draw[->] (phys.south) -- (lb) -| (reord.south);
  \node[font=\scriptsize] at ($(reord.south)!0.5!(phys.south)+(0,-1.18)$) {repeat until photon terminates};
  \draw[->,dashed] (reord.north) to[out=70,in=110] node[above=2pt,font=\scriptsize]{analytic navigation} (phys.north);
\end{tikzpicture}}
\caption{Data flow and per photon bounce loop of our GPU optical Monte Carlo (Simphony). Geant4 generates optical photon distributions on the host. \texttt{optixLaunch} starts one thread per photon. Each thread loops: \optixreorder{} (Section~\ref{sec:methods_ser}), then a BVH trace to the next boundary (the dominant ${\sim}70\%$ of kernel time), then the optical physics, repeating until the photon is absorbed, escapes or is detected at which point its hit is recorded. The dashed bypass is the newly introduced hybrid transport where the ray tracing is skipped (Section~\ref{sec:r2coh_hint}): deep inside the argon volume the next boundary is known analytically, so the BVH trace step is skipped while the physics still runs.}\label{fig:pipeline}
\end{figure}
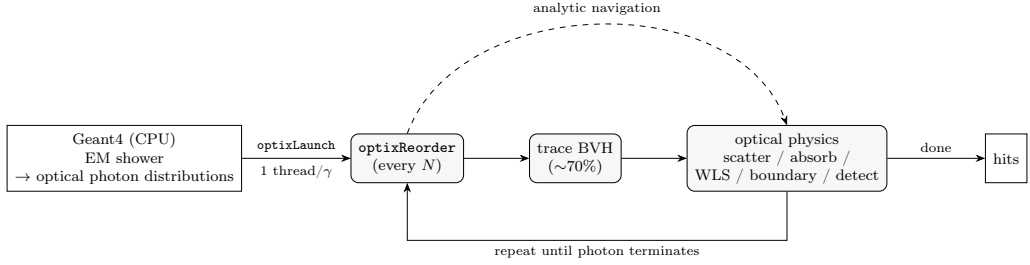

We use Simphony~\cite{companion}, built on Opticks~\cite{blyth_opticks}, CUDA and NVIDIA OptiX~9. Optical photon distributions emitted by Geant4 scintillation and Cherenkov processes are offloaded to the GPU, where a single \texttt{optixLaunch} propagation kernel transports the photons. Each thread executes a bounce loop: trace the geometry to the next boundary, apply the appropriate optical process, and repeat until the photon is detected, absorbed, or escapes. The overall data flow and the structure of the per photon bounce loop are shown in Fig.~\ref{fig:pipeline}. Within the GPU optical propagation step measured here, this propagation kernel accounts for \SI{99.5}{\percent} of GPU cycles. The remaining GPU work is bounding volume hierarchy (BVH) construction and hit compaction. 

\subsection{SIMT divergence and its metric}\label{sec:methods_divergence}

We quantify warp coherence using \warpeff{} as the metric: the average number of active, unmasked threads per issued warp instruction, out of the 32-thread warp width. A fully coherent kernel issues each instruction with all 32 lanes active. Execution divergence, whether caused by different branch paths or by different photon termination times, lowers this average because some lanes are inactive while the warp continues to issue instructions for the remaining lanes. We report the Nsight Compute counter \texttt{smsp\_\_thread\_inst\_executed\_per\_inst\_executed.ratio}, together with achieved occupancy and, separately, branch efficiency. Achieved occupancy is the number of warps resident on a multiprocessor relative to the maximum it can hold, it is a different quantity from \warpeff{}, a kernel can be highly occupied while most lanes within its issuing warps are inactive, and the two must not be confused. Branch efficiency measures whether branch targets are uniform across the warp. Throughout this paper, ``\warpeff{}'' denotes the active lane metric and ``execution divergence'' denotes the lane underutilisation it measures. The three SIMT divergence modes are compared in Listing~\ref{lst:divergence}. This paper shows that SER acts on the second.

\begin{lstlisting}[float=t,caption={The three SIMT divergence modes for a 32 lane warp. \emph{(1)} Branch divergence serialises control paths within a single instruction. \emph{(2)} Execution (lifetime) divergence is variation in the number of loop iterations: the warp stays resident until its slowest lane, and finished lanes sit idle. \emph{(3)} Memory divergence is scattered addressing. SER targets \emph{(2)} by repacking the surviving lanes. In our optical photon kernel, branch \emph{(1)} and memory \emph{(3)} are measured to be minor (Table~\ref{tab:expanded}) and \emph{(2)} dominates, because per photon bounce counts span orders of magnitude.},label=lst:divergence]
// (1) branch divergence: lanes take different control paths
if (cond[lane])  A();  else  B();        // warp executes A and B serially

// (2) execution / lifetime divergence: lanes iterate different counts   <-- SER target
for (i = 0; i < n[lane]; ++i)  step();   // warp runs to max(n); finished lanes sit idle

// (3) memory / data divergence: lanes touch scattered addresses
x = data[idx[lane]];                     // 1 load expands into many memory transactions
\end{lstlisting}

\subsection{Shader Execution Reordering}\label{sec:methods_ser}

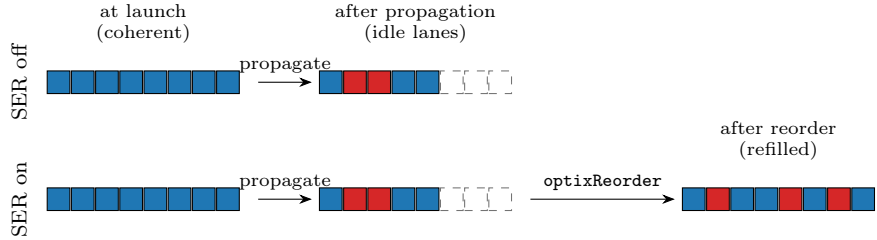
\begin{figure}[t]
\centering
\begin{tikzpicture}[font=\footnotesize,x=1cm,y=1cm,>={Stealth[]},
  act/.style={draw,fill=actcol,minimum width=3.0mm,minimum height=3.0mm,inner sep=0pt},
  actr/.style={draw,fill=redcol,minimum width=3.0mm,minimum height=3.0mm,inner sep=0pt},
  idle/.style={draw,dashed,draw=black!55,minimum width=3.0mm,minimum height=3.0mm,inner sep=0pt}]
  \definecolor{actcol}{RGB}{31,119,180}
  \definecolor{redcol}{RGB}{214,39,40}
  \def\yoff{0}\def\yon{-1.55}\def\s{0.32}\def\xa{0.5}\def\xb{4.1}\def\xc{8.9}
  \node[anchor=east] at (0.2,\yoff) {\rotatebox{90}{SER off}};
  \node[anchor=east] at (0.2,\yon) {\rotatebox{90}{SER on}};
  \foreach \i in {0,...,7}{\node[act] at (\xa+\i*\s,\yoff){};}
  \node[act] at (\xb+0*\s,\yoff){};\node[actr] at (\xb+1*\s,\yoff){};\node[actr] at (\xb+2*\s,\yoff){};\node[act] at (\xb+3*\s,\yoff){};\node[act] at (\xb+4*\s,\yoff){};
  \foreach \i in {5,6,7}{\node[idle] at (\xb+\i*\s,\yoff){};}
  \draw[->] (\xa+8*\s+0.08,\yoff) -- node[above,font=\scriptsize]{propagate} (\xb-0.25,\yoff);
  \foreach \i in {0,...,7}{\node[act] at (\xa+\i*\s,\yon){};}
  \node[act] at (\xb+0*\s,\yon){};\node[actr] at (\xb+1*\s,\yon){};\node[actr] at (\xb+2*\s,\yon){};\node[act] at (\xb+3*\s,\yon){};\node[act] at (\xb+4*\s,\yon){};
  \foreach \i in {5,6,7}{\node[idle] at (\xb+\i*\s,\yon){};}
  \draw[->] (\xa+8*\s+0.08,\yon) -- node[above,font=\scriptsize]{propagate} (\xb-0.25,\yon);
  \draw[->] (\xb+8*\s+0.08,\yon) -- node[above,font=\scriptsize]{\optixreorder} (\xc-0.25,\yon);
  \node[act] at (\xc+0*\s,\yon){};\node[actr] at (\xc+1*\s,\yon){};\node[act] at (\xc+2*\s,\yon){};\node[act] at (\xc+3*\s,\yon){};\node[actr] at (\xc+4*\s,\yon){};\node[act] at (\xc+5*\s,\yon){};\node[actr] at (\xc+6*\s,\yon){};\node[act] at (\xc+7*\s,\yon){};
  \node[anchor=south,font=\scriptsize,align=center] at (\xa+3.5*\s,0.4) {at launch\\(coherent)};
  \node[anchor=south,font=\scriptsize,align=center] at (\xb+3.5*\s,0.4) {after propagation\\(idle lanes)};
  \node[anchor=south,font=\scriptsize,align=center] at (\xc+3.5*\s,\yon+0.4) {after reorder\\(refilled)};
\end{tikzpicture}
\caption{The two execution models, following one warp (each square is one lane, eight of the 32 lanes are drawn, a solid lane is active, dashed is idle after its photon terminated. The two solid colours mark photons in different states of work). At launch every lane carries a fresh photon in the same state (one colour: coherent). As propagation proceeds the shorter lived photons terminate, leaving idle lanes, and the survivors drift into different work states (two colours), so the warp is both depleted and divergent. \emph{SER off} (top) leaves the warp in this state for the rest of its life. \emph{SER on} (bottom) calls \optixreorder{}, which consolidates the survivors of many such depleted warps back into full warps (and frees the emptied ones): every lane is active again. The refilled warp is still a mix of colours, SER recovers the idle lanes \emph{without} having to sort the work by type, consistent with the measurements of Section~\ref{sec:mechanism} identifying idle lane recovery.}\label{fig:execmodel}
\end{figure}

The \optixreorder{} device function invokes SER, allowing the OptiX runtime to regroup in flight threads before they continue execution (Fig.~\ref{fig:execmodel}). SER is hardware accelerated: from the Ada generation onward the regrouping of threads is performed by the GPU scheduling hardware together with the driver and OptiX runtime, not by an application level sort of ray state in global memory. This distinguishes it from software ray reordering schemes such as wavefront path tracing and explicit ray sorting (Section~\ref{sec:discussion}), which compact work between separate kernel launches. In practice, enabling SER changes more than a single instruction: the launch moves from a wide thread per photon grid to a launch configuration where SER is enabled (Section~\ref{sec:methods_benchmark}). Section~\ref{sec:mechanism} decouples the speed-up of these two contributions with a control build in which the reorder call is compiled in but never executes.

We invoke \optixreorder{} every $N$ bounces inside the bounce loop (Listing~\ref{lst:loop}), this single call is the only source level change to the kernel. The interval $N$ controls the tradeoff between the cost of reordering and the coherence restored by reordering. The relationship is shown in Section~\ref{sec:r2coh_interval}.

Listing~\ref{lst:loop} shows the production code path. \texttt{trace()} wraps the fused \texttt{optixTrace()} (traversal and shader invocation in one call). We do not use the split \texttt{optixTraverse} and \texttt{optixInvoke} calls. We reorder at the top of the loop rather than between traversal and shading because the divergence we target is lifetime divergence of the whole loop body: the regrouping must benefit the next trace \emph{and} the optical physics that follows it. In a megakernel both execute inside the ray generation program.

\begin{lstlisting}[float=t,caption={The \texttt{optixLaunch} propagation kernel, one thread per photon. Shader Execution Reordering is the single \texttt{optixReorder} call at the top of the loop; the trace and physics are unchanged. The second argument is the number of application coherence hint bits taken from the first.},label=lst:loop]
int bounce = 0;
while (bounce < maxBounce && p.time < maxTime) {
    if (bounce % N == 0)
        optixReorder(bounce / N, hintBits); // SER: regroup live threads into warps
    trace(handle, p.pos, p.mom, &prd);       // BVH traversal to next boundary 
    cmd = propagate(bounce, rng, p);         // bulk process / boundary / detect
    bounce++;
    if (cmd == BREAK) break;                 // photon absorbed, detected, or escaped
}
evt.photon[idx] = p;                         // write final photon state
\end{lstlisting} The call may also carry an application supplied coherence hint, used in Section~\ref{sec:r2coh_hint}. SER changes scheduling but not the photon physics. For every configuration reported here, the hit output is bit identical between paired SER off and SER on runs with the same seed.

\subsection{Benchmark and profiling}\label{sec:methods_benchmark}

\begin{figure}[t]
\centering
\includegraphics[width=0.50\linewidth]{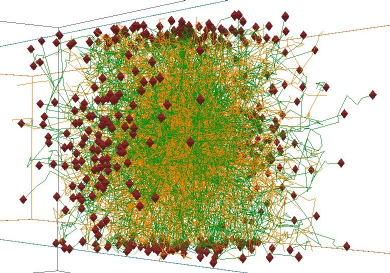}\hfill
\includegraphics[width=0.45\linewidth]{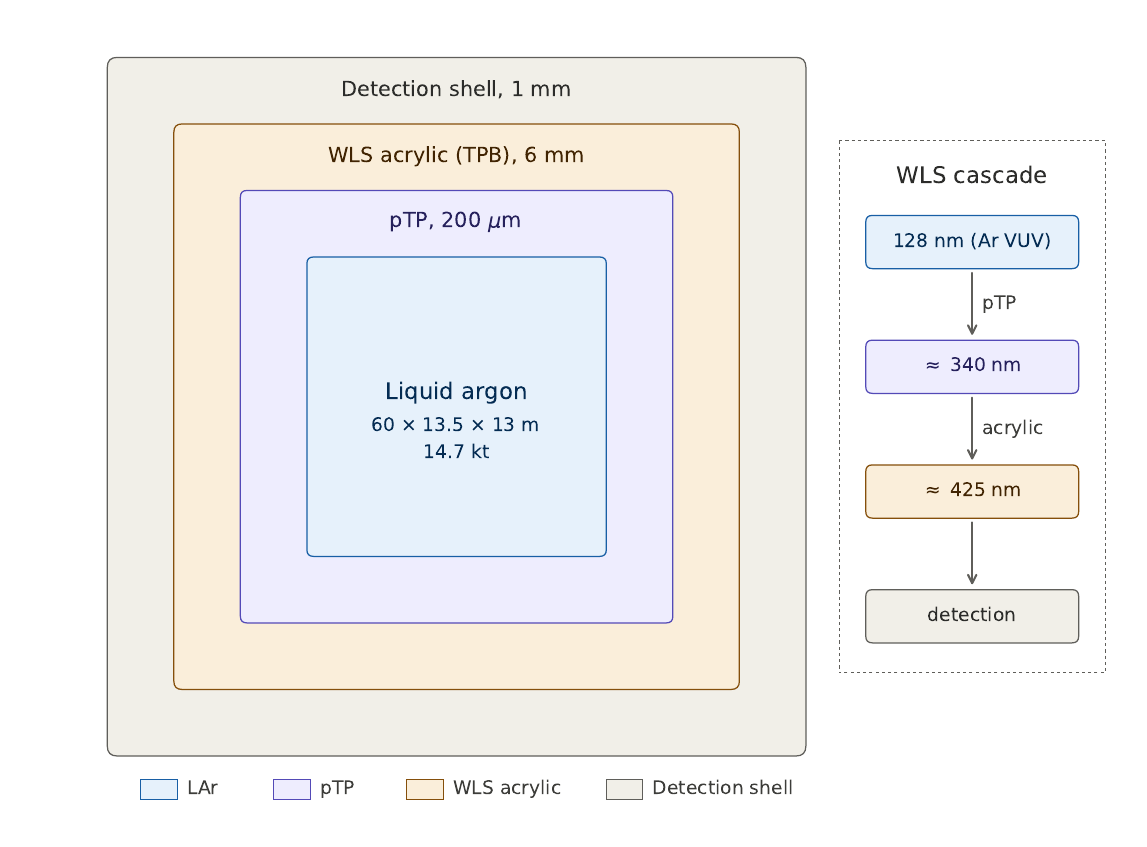}
\caption{The benchmark geometry and a representative photon load~\cite{companion}. \emph{Left:} optical photons from an electron event, propagated on the GPU (green, detected; orange, after WLS returned to the LAr volume and then reentered WLS and was detected). \emph{Right:} the simplified \SI{14.7}{\kilo\tonne} liquid argon TPC, a \SI{60}{\meter}\,$\times$\,\SI{13.5}{\meter}\,$\times$\,\SI{13}{\meter} argon volume nested in a \SI{200}{\micro\meter} \emph{p}-terphenyl layer, \SI{6}{\milli\meter} wavelength shifting acrylic, and a \SI{1}{\milli\meter} photon counting detector together with the two stage wavelength shifting cascade it implements (\SI{128}{\nano\meter} argon VUV $\to$ ${\approx}\SI{340}{\nano\meter}$ in pTP $\to$ ${\approx}\SI{425}{\nano\meter}$ in acrylic $\to$ detection).}\label{fig:geom}
\end{figure}

The benchmark geometry (Fig.~\ref{fig:geom}) is a simplified \SI{14.7}{\kilo\tonne} LArTPC: a \SI{60}{\meter}\,$\times$\,\SI{13.5}{\meter}\,$\times$\,\SI{13}{\meter} liquid argon volume wrapped in a wavelength shifting \SI{200}{\micro\meter} \emph{p}-terphenyl layer, \SI{6}{\milli\meter} of wavelength shifting acrylic, and a \SI{1}{\milli\meter} idealised photon counting detector~\cite{companion}. The test event consists of optical photons generated by a high energy electron: the primary is a \SI{2.5}{\giga\electronvolt} electron, representative of the electromagnetic shower from a charged current electron neutrino interaction in the LArTPC, whose scintillation and Cherenkov light amounts to ${\sim}61$ million optical photons per event. These photons are the workload propagated on the GPU.

All runs use an NVIDIA RTX~4090. Profiling uses Nsight Compute~2025.3 with kernel replay at locked base clock; production wall times are measured without the profiler at boost clock. Locking the clock to the base frequency makes the cycle count reproducible across Nsight's kernel replay passes. The wall time speedup, measured at boost clock without the profiler, is the most relevant metric for production. The SER comparison is obtained by compiling out \optixreorder{}. Every speedup in this paper is quoted relative to this SER off build: the same one thread per photon megakernel, geometry, seeds, and optical tables, with the single reorder call compiled out. The GPU baseline is itself almost two orders of magnitude faster than single-thread Geant4 optical transport on CPU, that comparison and the underlying physics validation are reported in another paper~\cite{companion}. That paper uses the SER speedup but not the analytical navigation one.

All measurements use a single RTX~4090 (Ada, \SI{24}{\giga\byte}) at its stock \SI{450}{\watt} power limit, with NVIDIA driver 580.105.08, CUDA~13.0, and OptiX~9.0. The kernel is compiled with \texttt{nvcc -O3} (C++17, no fast math flags) to PTX against a compute capability 7.5 baseline and JIT compiled by the driver for the Ada device. Nsight Compute's default clock control pins the core clock to its base frequency during every replay pass. Production wall times use the driver's default boost behaviour with no application clock pinning, and the quoted uncertainties are the spread over repeated launches. Random numbers are produced by cuRAND Philox4x32-10 generator, one subsequence per photon selected by its global photon index, which is what makes the hit output independent of thread scheduling and hence of reordering.

\section{SER speedup and its mechanism}\label{sec:mechanism}

SER execution reduces the GPU cycle count of the propagation kernel by a factor of \num{16.55(23)}, reordering every fourth bounce, relative to the same one thread per photon OptiX megakernel with the single reorder call compiled out. The simulation output is bit identical. Independent profiler free measurements of the whole optical photon simulation, comprising the host to device transfer of photon distributions, GPU initialization, the propagation kernel, and copying the hits back to the host yield a \num{14.72(30)}$\times$ wall time speedup. This is slightly lower than the kernel cycle ratio because these host and transfer overheads, approximately \SI{120}{\milli\second} per event, are not accelerated by SER. The optical photon simulation of the benchmark event takes \SI{14.29(36)}{\second} without SER and \SI{0.99(4)}{\second} with the reorder executing, falling to \SI{0.87(6)}{\second} with the analytic navigation additions of Section~\ref{sec:r2coh_hint} (Table~\ref{tab:ladder}), the Geant4 EM shower generation on the host is excluded, being identical in all configurations.

Enabling SER changes the compiled pipeline and launch configuration, while executing \optixreorder{} changes the runtime behaviour by reordering threads. To decouple the effect of both on the speedup, we compared four builds (Table~\ref{tab:ladder}) on the same geometry and seeds, with bit identical hit output across all of them. Build (b) runs $3.17\times$ faster than (a) in kernel cycles ($3.09\times$ in wall time): the pipeline and launch configuration changes alone recover part of the drained warp cost even without any reordering. Build (c) runs a further $5.10\times$ faster in cycles ($4.69\times$ wall): the reorder actually executing yields the additional speed-up. A further test replacing the production key with a constant, non-informative value produced results consistent with build (c) within uncertainties in two independent measurement campaigns, indicating that this gain comes from executing the reorder rather than from the key content. Build (d) adds a further $1.17\times$ in cycles ($1.13\times$ wall), described in Section~\ref{sec:r2coh_hint}. The speed-up decomposes as follows: $3.17 \times 5.10 = 16.1\times$ in cycles and $3.09 \times 4.69 = 14.5\times$ in wall time.

\begin{table}[t]
\caption{The four control builds decomposing the speed-up due to the pipeline and launch configuration changes associated with enabling SER and to executing the reorder. Kernel cycles are Nsight Compute elapsed GPC cycles at locked base clock. The cycle speedups decompose as follows: $3.17 \times 5.10 = 16.1\times$. Wall time is the whole GPU optical photon simulation per launch (host to device transfer, GPU initialisation, propagation kernel, and hit copy back) at boost clock without the profiler, where the decomposition is $3.09 \times 4.69 = 14.5\times$. The wall ratios sit below the cycle ratios because the fixed host and transfer overheads are not accelerated. Build (d) adds the analytic navigation of Section~\ref{sec:r2coh_hint} with its in volume coherence hint, a further $1.17\times$ in cycles and $1.13\times$ in wall time over (c). The full hardware counter comparison exists for builds (a) and (c) (Table~\ref{tab:expanded}).}

\begin{tabular}{@{}cp{4.4cm}rrrr@{}}
\toprule
 & & cycles & wall & \multicolumn{2}{c@{}}{speedup ($\times$)} \\
\cmidrule(l){5-6}
build & & ($10^{9}$) & (s) & kernel & wall \\
\midrule
(a) & \raggedright SER off & \num{38.2(10)} & \num{14.29(36)} & 1 & 1 \\
(b) & \raggedright SER on, never executes & \num{12.08(34)} & \num{4.62(13)} & \num{3.17(1)} & \num{3.09(1)} \\
(c) & \raggedright reorder executed every 4th bounce & \num{2.37(12)} & \num{0.99(4)} & \num{16.1(4)} & \num{14.5(3)} \\
(d) & \raggedright executed, volume hint $+$ analytic navigation & \num{2.03(12)} & \num{0.87(6)} & \num{18.9(6)} & \num{16.4(7)} \\
\botrule
\end{tabular}
\label{tab:ladder}
\end{table}

\begin{figure}[t]
\centering
\includegraphics[width=0.92\linewidth]{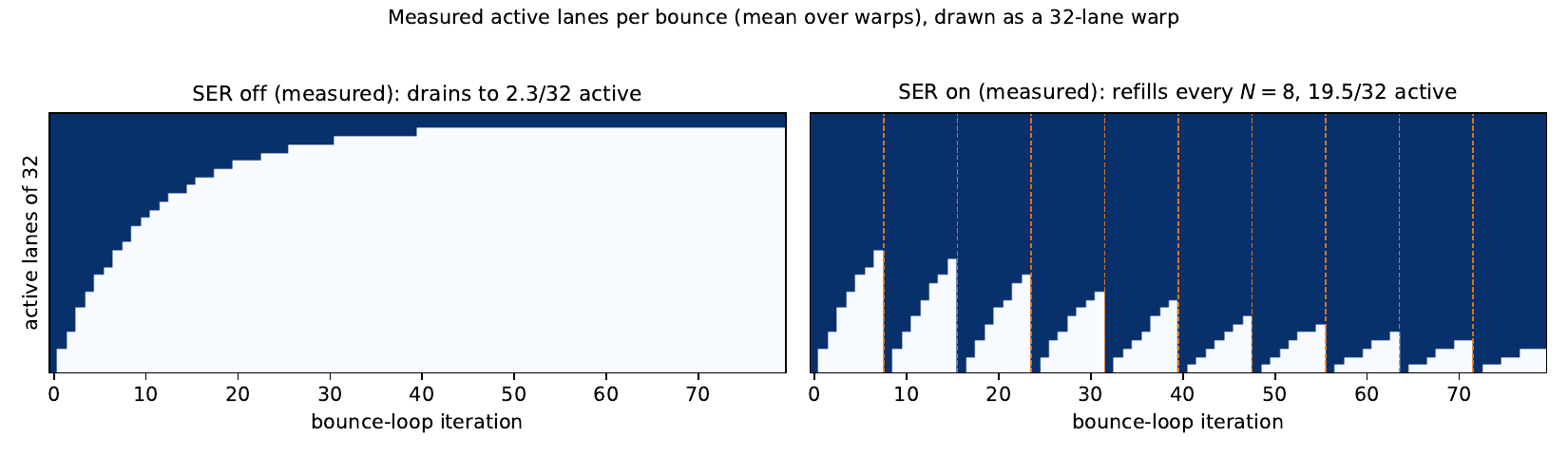}
\caption{Measured active lanes per bounce, drawn on the 32 lane warp. \texttt{\_\_activemask} instrumentation is used to measure this. \emph{Left} (off): the active lane count drains monotonically as photons terminate, leaving most lanes idle while a few long lived stragglers keep the warp resident, a loop mean of \num{2.3} of 32. \emph{Right} (on): a few lanes terminate between reorders, but each reorder (orange ticks, every $N=8$) refills the warp toward full by consolidating survivors from many depleted warps, so the active lane count saws back to full rather than decaying away, a loop mean of \num{19.5} of 32. These loop only means sit just below the whole kernel Nsight values of Table~\ref{tab:expanded} ($2.64$ and $20.4$), which also average over the code outside the bounce loop: the photon setup before the first bounce and the final state write after the loop execute with all 32 lanes active, pulling the whole kernel mean slightly above the loop only mean.}\label{fig:straggler}
\end{figure}

Figure~\ref{fig:straggler} shows the measured per bounce active lane count as a 32 lane warp heatmap, from in kernel \texttt{\_\_activemask} instrumentation (one sample per warp per bounce, post reorder). Without reordering, most lanes terminate early and remain inactive until the longest lived photons complete, so the warp drains. With periodic reordering, surviving photons are consolidated into warps with more active lanes every $N$ bounces, so the active lane count refills toward full at each reorder and the active lane fraction stays high.

\subsection{Execution divergence}\label{sec:mechanism_decomp}

Table~\ref{tab:expanded} reports the Nsight Compute profiler results for  SER off versus on settings on the \texttt{optixLaunch} propagation kernel. \Warpeff{} rises by $7.72\times$, from \num{2.64} to \num{20.37} active lanes of 32, corresponding to \SI{8.3}{\percent} and \SI{63.7}{\percent} of the warp width. This active lane recovery is the one mechanism confirmed independently of the profiler's launch geometry (Section~\ref{sec:methods_benchmark}): it is reproduced inside the running kernel by the \texttt{\_\_activemask} measurement of Fig.~\ref{fig:straggler} (\num{2.3}\,$\to$\,\num{19.5} lanes). Taken together the counters are consistent with the cycle ratio: warp instructions issued fall by \num{8.7}$\times$ (the \num{1.13}$\times$ fewer predicated instructions times the \num{7.72}$\times$ recovered lanes), issue slot activity rises \num{1.77}$\times$ (latency hiding from the higher occupancy), and the product of the two, \num{15.4}$\times$, agrees with the measured \num{16.3(5)}$\times$ cycle reduction to within ${\sim}6\%$. Occupancy and issue slot activity also depend on the launch configuration, which SER itself changes, but we quote them as the best measures the profiler offers. Active lane recovery is therefore the largest measured contribution, with the instruction count and issue rate changes covering the remainder.

\begin{figure}[h!]
\centering
\includegraphics[width=0.92\linewidth]{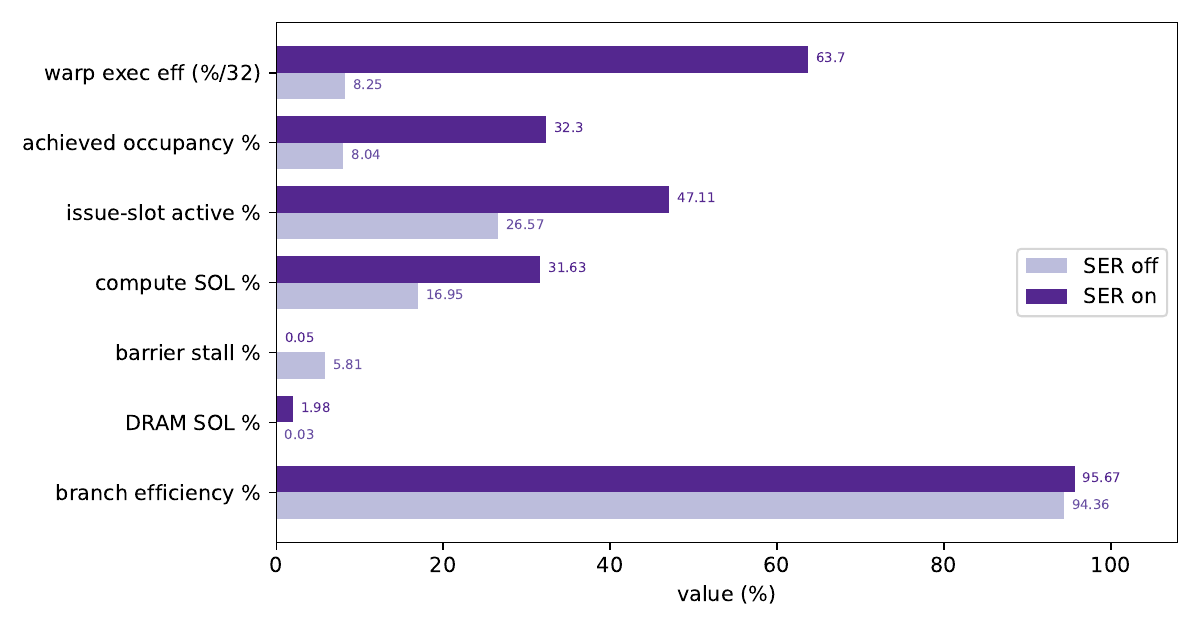}
\caption{Per metric, SER off versus on. Branch efficiency and global load sectors per request change little. Barrier stalls decrease, while issue slot activity and warp execution efficiency increase. These changes are consistent with reduced execution divergence.}\label{fig:stallbar}
\end{figure}

Branch efficiency changes only from \SI{94.36}{\percent} to \SI{95.67}{\percent}; branch nonuniformity is therefore approximately \SI{5}{\percent} in both builds. That small change cannot plausibly explain an order of magnitude cycle reduction. Likewise, global load sectors per request change from \num{1.00} to \num{1.15}. DRAM throughput remains low after SER, at \SI{1.98}{\percent} of peak, and the measured L2 hit rate is \SI{99.9}{\percent}; these place the SER on kernel in a latency- and occupancy bound regime, not a bandwidth bound one. The stall composition is consistent with the same interpretation (Fig.~\ref{fig:stallbar}): barrier stalls fall from \SI{5.81}{\percent} to \SI{0.05}{\percent}. These counters support the conclusion that the dominant removable cost is execution divergence from inactive lanes.

\begin{table}[t]
\caption{Nsight Compute profile of the \texttt{optixLaunch} propagation kernel, SER off vs.\ on. The largest change is \warpeff{}, branch efficiency moves little, and the memory counters show no sign of a bandwidth limit (Section~\ref{sec:mechanism_decomp}). The SER on build is profiled under a cooperative launch geometry (Section~\ref{sec:methods_benchmark}): the geometry independent rows (elapsed cycles, \warpeff{}) are directly comparable, and the \warpeff{} change is confirmed in kernel (Fig.~\ref{fig:straggler}).}\label{tab:expanded}
\centering
\resizebox{\linewidth}{!}{%
\begin{tabular}{@{}lrrr@{}}
\toprule
Metric & SER off & SER on & ratio \\
\midrule
\warpeff{} (lanes/inst, max 32) & \num{2.640(26)} & \num{20.377(45)} & $7.72\times$ \\
Achieved occupancy (\%) & \num{8.02(1)} & \num{32.29(3)} & --- \\
Elapsed GPC cycles & \num{3.75(18)e10} & \num{2.30(8)e9} & $(16.3\pm0.5)\times$ \\
Branch efficiency (\%) & \num{94.36} & \num{95.67(3)} & --- \\
Global LD sectors/request & \num{1.00} & \num{1.15(1)} & --- \\
DRAM throughput (\% peak) & \num{0.03} & \num{1.98(4)} & --- \\
Stall, barrier (\%) & \num{5.81(4)} & \num{0.05} & --- \\
Issue slot active (\%) & \num{26.57(1)} & \num{47.1(3)} & $1.77\times$ \\
\botrule
\end{tabular}}
\end{table}

\newpage

\section{Generalization of the method: using one divergence source}\label{sec:scaling_control}

In order to decouple the contributions of transport in the WLS layers and bulk scattering in LAr in the Monte Carlo simulation, in the following dedicated study we removed the wavelength shifting and acrylic shells, wrapping the argon bulk directly in the photon detector and leaving the argon optical properties unchanged. In this scenario the photons perform a random walk in the argon bulk by repeated Rayleigh scattering until they are absorbed or reach the detector, so the remaining source of execution divergence is the spread in photon lifetimes, the number of scatters before termination, not trapping in the shells. This modified geometry study isolates bulk scattering lifetime variation from trapping and shows that bulk transport alone still produces a SER benefit in the heavily scattering regime.

\begin{figure}[t]
\centering
\includegraphics[width=\linewidth]{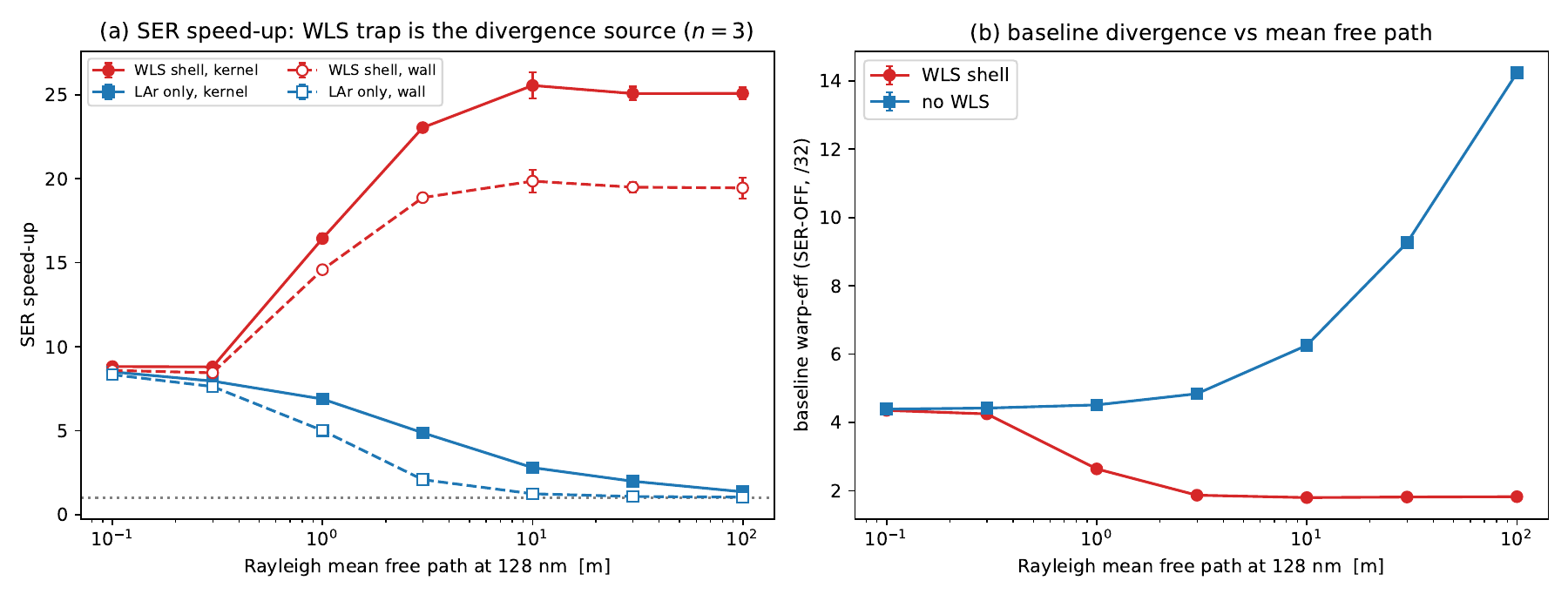}
\caption{Geometry comparison: SER speedup (left; solid\,$=$\,locked clock kernel cycles, dashed\,$=$\,end-to-end optical photon simulation wall time) and baseline \warpeff{} (right) vs.\ scattering length, with the wavelength shifting shell (red) and for argon\,$+$\,detector only (blue). Removing the shell collapses the high divergence speedup and inverts the trend. The wall speedup tracks the kernel speedup but saturates below it at large mean free path, the WLS shell wall gain plateaus near \num{20}$\times$ while the kernel reaches \num{25}$\times$ because the fixed host and transfer tail of the optical photon simulation does not scale. }\label{fig:nowls}
\end{figure}

\begin{figure}[t]
\centering
\includegraphics[width=\linewidth]{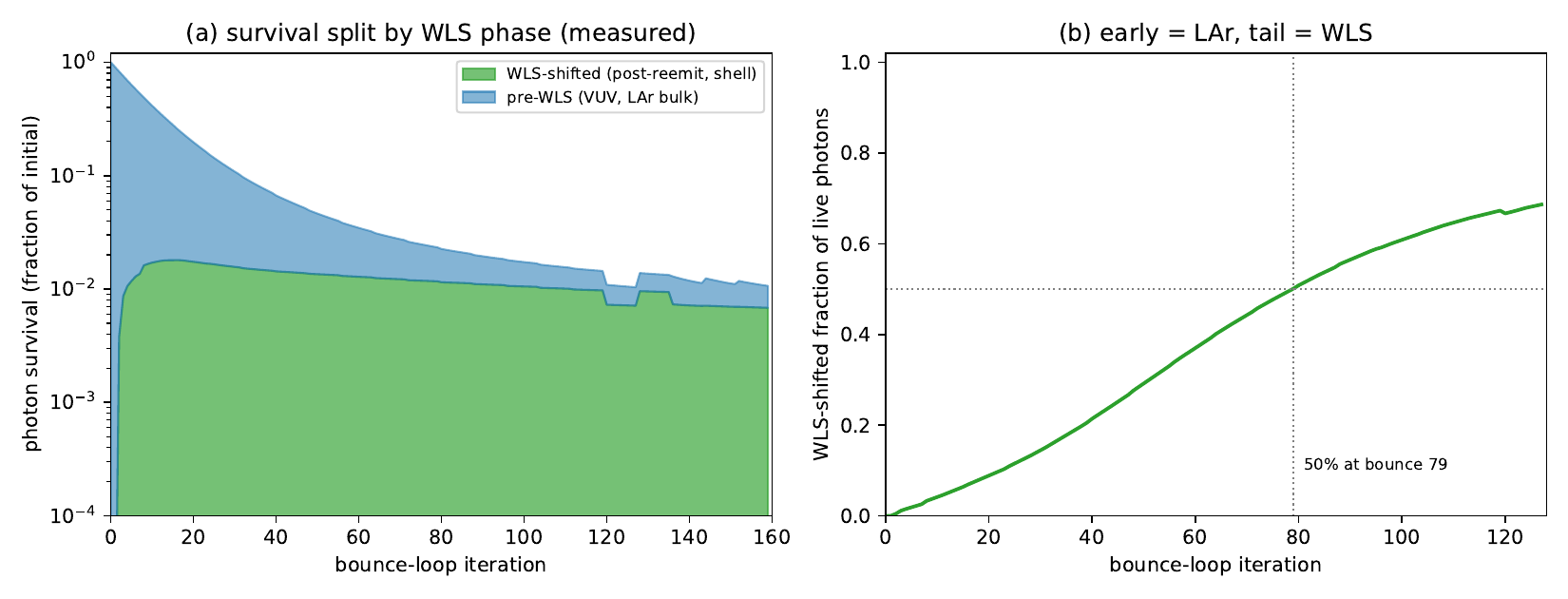}
\caption{Divergence source identified from within the test geometry. Each photon's wavelength shifting state (\texttt{BULK\_REEMIT}) is recorded per bounce. \emph{(a)} Surviving photon fraction split into preshift (VUV, argon bulk; blue) and postshift (reemitted, shell; green): the preshift population decays within tens of bounces while the postshift population persists as a heavy tail. \emph{(b)} Wavelength shifted fraction of surviving photons vs.\ bounce, crossing \SI{50}{\percent} at bounce 79. The heavy tailed stragglers that keep warps resident are therefore predominantly shell trapped wavelength shifted photons, with a secondary contribution from long argon random walks.}\label{fig:wlsphase}
\end{figure}

Figure~\ref{fig:nowls} compares the two geometries. With the WLS shell the SER speedup rises with scattering length, reaching ${\sim}\num{25}\times$ at long mean free path; without the shell it falls instead: at long mean free path a photon leaves the argon in approximately one step, so the SER off baseline is already much more coherent, with \warpeff{} reaching \num{14.2} active lanes of 32. In the long-mean-free-path regime, where the WLS-shell geometry exhibits high baseline divergence, removing the shell collapses the speedup from approximately \num{25}$\times$ to \num{1.4}$\times$. The two geometries converge only in the heavily scattering limit, where photons are absorbed in the bulk before reaching the shell system. The same geometry comparison, measured in end-to-end optical photon simulation wall time (dashed curves in Fig.~\ref{fig:nowls}) reproduces the identical pattern at a lower magnitude: the shell wall speedup plateaus near \num{20}$\times$ where the kernel reaches \num{25}$\times$, the difference being the fixed host and transfer tail of the optical photon simulation, while the no shell geometry collapses to ${\sim}\num{1}\times$. These observations identify trapping in the WLS layers as the dominant source of the heavy tailed lifetime distribution in this kiloton scale benchmark. The per bounce composition of the surviving population in the unmodified
geometry confirms this (Fig.~\ref{fig:wlsphase}): the wavelength shifted
fraction of survivors crosses \SI{50}{\percent} at bounce 79 and dominates the tail.

\newpage

\section{Tuning the reorder: interval and coherence hint}\label{sec:r2coh}

\subsection{Reorder interval}\label{sec:r2coh_interval}

SER is invoked every $N$ bounces, trading the overhead of the reorder against the coherence it restores. Table~\ref{tab:interval} shows both the kernel and the wall speedup for different $N$. \Warpeff{} stays high and nearly flat for small $N$, and both the kernel cycle count and the wall time are minimised at $N=4$, where the kernel speedup peaks at \num{16.55(23)}$\times$ and the wall speedup at \num{14.72(30)}$\times$. Reordering too rarely allows warp coherence to degrade: by $N=16$, \warpeff{} has dropped to \num{17.6} of 32 and the speedup falls to \num{15.30(11)}$\times$ kernel and \num{13.68(15)}$\times$ wall. 

\begin{table}[h!]
\caption{Reorder interval scan: locked clock kernel cycle speedup and end-to-end optical photon simulation wall speedup vs.\ reordering every $N$th bounce. Both peak at $N=4$, which balances reorder overhead against coherence decay. Reordering every bounce ($N=1$) and reordering rarely ($N=16$) are the slowest settings but still exceed \num{13.6}$\times$ wall.}\label{tab:interval}
\begin{tabular}{@{}rrrrr@{}}
\toprule
$N$ & \warpeff{} & cycles ($10^{9}$) & kernel speedup ($\times$) & wall speedup ($\times$) \\
\midrule
1 & 21.5 & 2.502 & $15.24\pm0.47$ & $13.68\pm0.35$ \\
2 & 21.9 & 2.349 & $16.22\pm0.31$ & $14.59\pm0.33$ \\
\textbf{4} & \textbf{22.0} & \textbf{2.301} & $\mathbf{16.55\pm0.23}$ & $\mathbf{14.72\pm0.30}$ \\
5 & 21.6 & 2.319 & $16.43\pm0.30$ & $14.66\pm0.22$ \\
8 & 20.4 & 2.353 & $16.19\pm0.20$ & $14.69\pm0.15$ \\
10 & 19.6 & 2.368 & $16.08\pm0.11$ & $14.39\pm0.15$ \\
16 & 17.6 & 2.488 & $15.30\pm0.11$ & $13.68\pm0.15$ \\
\botrule
\end{tabular}
\end{table}

\subsection{Region coherent hint for Shader Execution Reorder}\label{sec:r2coh_hint}

\optixreorder{} also accepts an application defined coherence hint. We use this hint to preserve coherence across an analytic empty space optimisation. Deep inside the large argon volume, the next possible forward boundary is the vessel wall and can be computed analytically, given its box shape. We do not trace the BVH deep inside the volume because we already know the photon cannot reach another volume before the wall. In such regions the BVH trace step, which accounts for approximately \SI{70}{\percent} of the propagation kernel time, can in principle be skipped (the dashed bypass in Fig.~\ref{fig:pipeline}):  the distance to the next boundary is computed analytically instead of by BVH traversal, while the optical physics is unchanged. When the fast path is taken it fills exactly the fields the BVH intersection would have returned: the distance to the wall from an analytic box intersection against the vessel faces, the outward normal of the exit face, and the boundary identity of the wall, so the downstream physics is identical on both paths. The sampled scattering and absorption distances still compete against the boundary distance, so a photon that interacts before the wall never reaches it in that specific step. Additionally, any direction change ends the advance and the next step recomputes the boundary distance along the new direction, the same per step guarantee Geant4 provides by limiting each straight substep to the boundary along the current direction. The hit output with this hybrid mode and full BVH traversal was measured to be bit identical. The analytic navigation is enabled only if the photon position is inside the liquid argon volume, which does not contain any daughter volumes. Grazing and near wall cases, where the analytic distance does not exceed the propagation epsilon, abandon the fast path and run the real BVH trace.

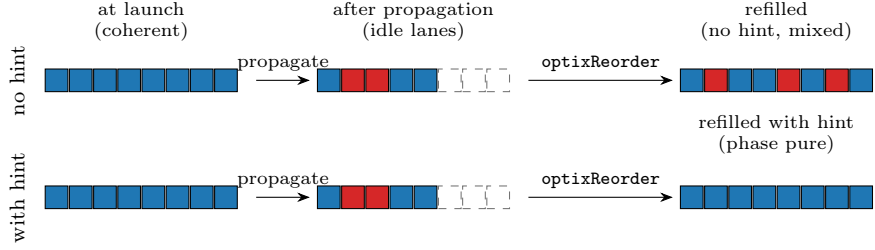
\begin{figure}[t]
\centering
\begin{tikzpicture}[font=\footnotesize,x=1cm,y=1cm,>={Stealth[]},
  act/.style={draw,fill=actcol,minimum width=3.0mm,minimum height=3.0mm,inner sep=0pt},
  actr/.style={draw,fill=redcol,minimum width=3.0mm,minimum height=3.0mm,inner sep=0pt},
  idle/.style={draw,dashed,draw=black!55,minimum width=3.0mm,minimum height=3.0mm,inner sep=0pt}]
  \definecolor{actcol}{RGB}{31,119,180}
  \definecolor{redcol}{RGB}{214,39,40}
  \def\yoff{0}\def\yon{-1.55}\def\s{0.32}\def\xa{0.5}\def\xb{4.1}\def\xc{8.9}
  \node[anchor=east] at (0.2,\yoff) {\rotatebox{90}{no hint}};
  \node[anchor=east] at (0.2,\yon) {\rotatebox{90}{with hint}};
  \foreach \i in {0,...,7}{\node[act] at (\xa+\i*\s,\yoff){};}
  \node[act] at (\xb+0*\s,\yoff){};\node[actr] at (\xb+1*\s,\yoff){};\node[actr] at (\xb+2*\s,\yoff){};\node[act] at (\xb+3*\s,\yoff){};\node[act] at (\xb+4*\s,\yoff){};
  \foreach \i in {5,6,7}{\node[idle] at (\xb+\i*\s,\yoff){};}
  \draw[->] (\xa+8*\s+0.08,\yoff) -- node[above,font=\scriptsize]{propagate} (\xb-0.25,\yoff);
  \draw[->] (\xb+8*\s+0.08,\yoff) -- node[above,font=\scriptsize]{\optixreorder} (\xc-0.25,\yoff);
  \node[act] at (\xc+0*\s,\yoff){};\node[actr] at (\xc+1*\s,\yoff){};\node[act] at (\xc+2*\s,\yoff){};\node[act] at (\xc+3*\s,\yoff){};\node[actr] at (\xc+4*\s,\yoff){};\node[act] at (\xc+5*\s,\yoff){};\node[actr] at (\xc+6*\s,\yoff){};\node[act] at (\xc+7*\s,\yoff){};
  \foreach \i in {0,...,7}{\node[act] at (\xa+\i*\s,\yon){};}
  \node[act] at (\xb+0*\s,\yon){};\node[actr] at (\xb+1*\s,\yon){};\node[actr] at (\xb+2*\s,\yon){};\node[act] at (\xb+3*\s,\yon){};\node[act] at (\xb+4*\s,\yon){};
  \foreach \i in {5,6,7}{\node[idle] at (\xb+\i*\s,\yon){};}
  \draw[->] (\xa+8*\s+0.08,\yon) -- node[above,font=\scriptsize]{propagate} (\xb-0.25,\yon);
  \draw[->] (\xb+8*\s+0.08,\yon) -- node[above,font=\scriptsize]{\optixreorder} (\xc-0.25,\yon);
  \foreach \i in {0,...,7}{\node[act] at (\xc+\i*\s,\yon){};}
  \node[anchor=south,font=\scriptsize,align=center] at (\xa+3.5*\s,0.4) {at launch\\(coherent)};
  \node[anchor=south,font=\scriptsize,align=center] at (\xb+3.5*\s,0.4) {after propagation\\(idle lanes)};
  \node[anchor=south,font=\scriptsize,align=center] at (\xc+3.5*\s,0.4) {refilled\\(no hint, mixed)};
  \node[anchor=south,font=\scriptsize,align=center] at (\xc+3.5*\s,\yon+0.45) {refilled with hint\\(phase pure)};
\end{tikzpicture}
\caption{The region hint in the execution model picture of Fig.~\ref{fig:execmodel}; both rows are SER on. The top does not pass the region hint to SER, the bottom does. Colours mark the volume each surviving photon occupies, blue is argon bulk, red is the wavelength shifting shell. \emph{No hint} (top): the default reorder carries no informative key, so it refills the idle lanes but leaves the refilled warp a mix of both volumes. \emph{With hint} (bottom): feeding the current volume into the coherence key makes groups surviving threads by current volume, so nearly all refilled warps contain lanes from a single region. This per warp volume coherence is what keeps the analytic fast path coherent under reordering and enables the additional \num{1.13}$\times$ speedup of Section~\ref{sec:r2coh_hint}.}\label{fig:hintexec}
\end{figure}

\begin{figure}[h!]
\centering
\includegraphics[width=\linewidth]{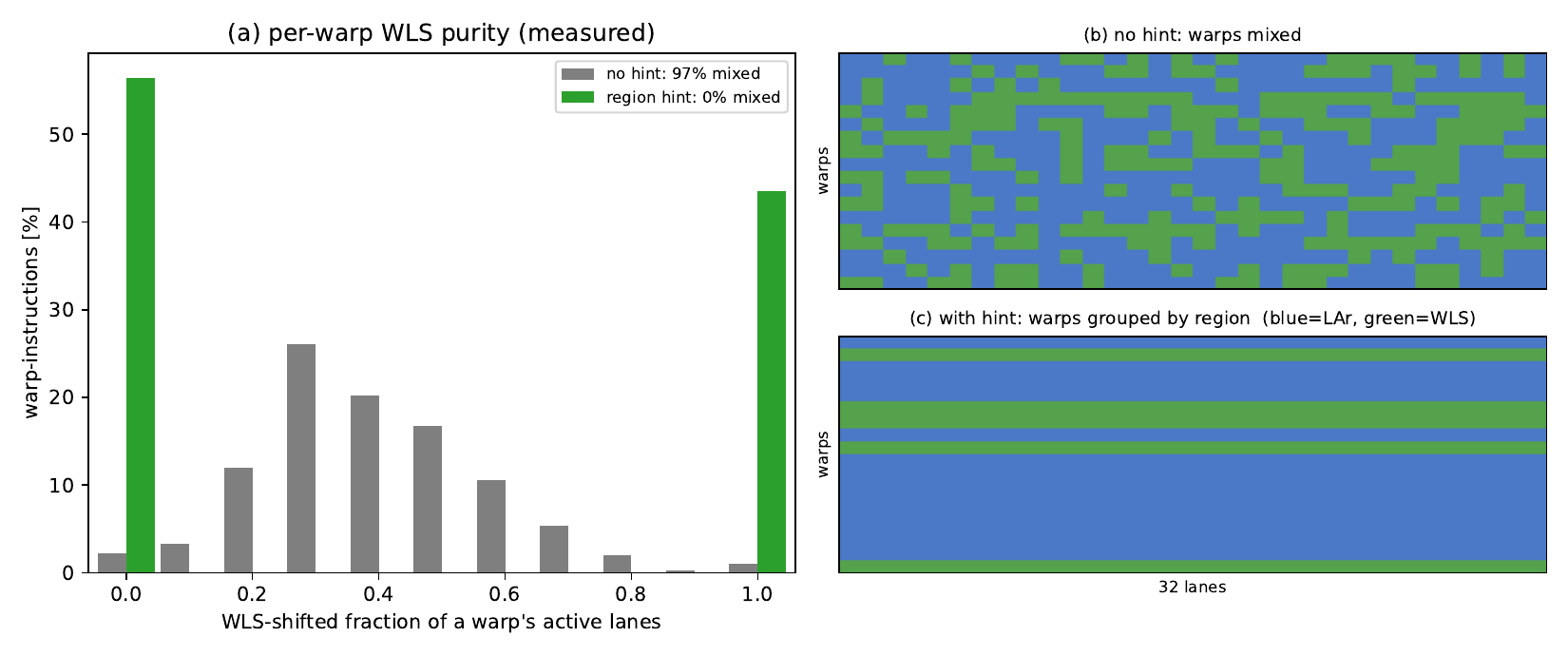}
\caption{The region hint regroups threads by volume. \emph{(a)} Per warp distribution of the wavelength shifted (\texttt{BULK\_REEMIT}) fraction of active lanes, sampled once per warp after each reorder: without an informative hint \SI{96.7}{\percent} of warps are mixed; with the region hint warps are phase pure (\SI{0.02}{\percent} mixed, bimodal at 0 and 1). \emph{(b,c)} Illustrative warps drawn from the measured distributions: without the hint lanes are mixed (b); with the hint warps are coherent bands of argon bulk (blue) and shell (green) photons (c). This grouping separates the analytic and BVH paths.}\label{fig:purity}
\end{figure}

A naive application of this method breaks even in this benchmark. Photons that do not invoke ray tracing are mixed in a warp with ray-traced photons, introducing branch divergence. Supplying a one bit in volume region hint makes the grouping coherent for both the ray-traced and non-ray-traced paths (Fig.~\ref{fig:hintexec}). With this hint, the analytical navigation optimisation gives a measured \num{1.13}$\times$ additional wall time speedup (\num{1.17}$\times$ in kernel cycles) on top of the SER baseline, while preserving bit identical results. Figure~\ref{fig:purity} shows this directly. We record each photon's wavelength shifting state after each reorder: each sample is one warp at one post reorder bounce, and a sample counts as mixed if the wavelength shifted fraction of its active lanes lies strictly between zero and one.  \SI{96.7}{\percent} of warp samples remain mixed in wavelength shifting phase, whereas with the region hint warps become almost phase pure, all argon bulk or all shell (only \SI{0.02}{\percent} of samples are mixed).

\newpage

\section{Discussion}\label{sec:discussion}

SER is most effective when the workload has a broad, heavy tailed per photon lifetime distribution. Here the tail comes primarily from WLS trapping, while bulk argon scattering alone still provides enough lifetime variation to yield a $7.0\times$ speedup in the modified geometry at the physically representative Rayleigh scattering length used in the benchmark. The magnitude of the gain depends on the baseline \warpeff{} and the weight of the lifetime tail. The method, however, is not specific to noble liquids. NVIDIA's own SER guidance recommends encoding loop termination state in the coherence hint for multibounce path tracers~\cite{ser_whitepaper}. A directly analogous example is mesh based Monte Carlo photon transport in biological tissue, implemented as an OptiX megakernel and reported to have strongly variable per photon path lengths~\cite{rtmmc}; reverberant acoustic ray tracing is a further candidate. GPU neutron transport Monte Carlo is a related case with documented divergence on both axes: per neutron collision counts before capture, fission, or leakage are heavy tailed (with a tail beyond 400 collisions~\cite{hamilton2019shift}) and the reaction sampled at each collision makes the physics control flow divergent. The neutron community has reordered in software for a decade. WARP radix sorts live neutrons by reaction type after every transport iteration~\cite{bergmann2015warp}, event based OpenMC sorts particles by material and energy to run roughly $6\times$ faster than history based execution on GPU~\cite{tramm2024openmc}, and Shift reports $2.6$--$5.6\times$ from event based restructuring, driven largely by recovered occupancy~\cite{hamilton2019shift}. They are conceptually similar: both group work by a coherence key. \optixreorder{} can only regroup threads inside an OptiX raygen megakernel, in our optical kernel the physics executes inside that megakernel alongside the trace (Listing~\ref{lst:loop}), so a single reorder call regroups the threads for both the trace and the subsequent physics. Production GPU neutron codes to our knowledge instead run their divergent cross section and reaction sampling in separate CUDA kernels, using ray tracing, if at all, only for geometry~\cite{salmon2019openmc}. Applying SER would therefore require restructuring the transport into a single OptiX raygen megakernel. Moreover, much of their event based gain comes from recovered occupancy, not from lane divergence alone~\cite{hamilton2019shift}, which is the component SER addresses.

The same divergence is independently quantified in the IceCube neutrino observatory's glacial-ice propagation, where it is met in software rather than in hardware. Its production codes report an average GPU load of \num{24.7} of 32 for the PPC code and \num{13.7} of 32 for CLSim, the latter rising to \num{17.0} of 32 after a dedicated optimization~\cite{chirkin_ppc,schwanekamp2022}. The underlying effect is execution divergence as well, a thread idling once its photon is absorbed. In IceCube each warp holds a small shared memory queue of photons and a lane reloads the next one when its own terminates, so finished lanes refill instead of idling and a warp drains only when its queue is empty. The CLSim optimization additionally removed a particularly expensive branch ~\cite{schwanekamp2022}. Our one-thread-per-photon baseline carries no such software refill, \num{2.64} of 32, because we leave the consolidation of survivors to reordering rather than to a hand tuned queue. \optixreorder{} then restores it to \num{20.4} of 32 from a single ray-generation call, and it can regroup survivors across warps rather than only within one. IceCube runs CUDA in analytic ice, not an OptiX pipeline, so applying SER there would require an OptiX port. The comparison confirms that warp divergence is the recognized efficiency bottleneck in GPU photon transport, addressed by IceCube in software and in this paper in hardware.

The hybrid pattern of Section~\ref{sec:r2coh_hint} maps directly onto IceCube: PPC already propagates analytically and culls detector modules with a software grid~\cite{chirkin_ppc}, and an OptiX port that instead traced a BVH on every step ran at roughly half its speed~\cite{schwanekamp2022}. One untested option is to host the analytic kernel in an OptiX ray-generation program and use hint only \optixreorder{} in place of the software photon pool. BVH traversal would then be used only for segments near a detector module, with analytic transport elsewhere. One design rule applies: the near module bit must enter the coherence hint, or the fast path breaks even (Section~\ref{sec:r2coh_hint}).


Reordering also reduces energy per event. Sampling board power with \texttt{nvidia-smi} (NVML) at \SI{100}{\milli\second} resolution during the propagation (three seeds at the headline \SI{2.5}{\giga\electronvolt}, no profiler, boost clock) and integrating it over the GPU active window (utilisation above \SI{20}{\percent}, which is somewhat broader than the timed optical photon simulation because it also covers BVH build and transfer ramp), SER shortens that window from \SI{14.2}{\second} to \SI{1.3}{\second} while the window averaged board power rises from \SI{158}{\watt} to \SI{184}{\watt}, the higher occupancy keeping more of the device busy. The net GPU energy of the optical propagation falls from \SI{2246(110)}{\joule} to \SI{237(17)}{\joule} per event, a $9.5\times$ reduction. The energy gain is smaller than the time gain precisely because the more efficient kernel draws more instantaneous power, but it remains close to an order of magnitude.



\section{Conclusions}\label{sec:conclusions}

We have shown that Shader Execution Reordering speeds up optical GPU Monte Carlo simulation of an electromagnetic shower representative of a charged current electron neutrino interaction by more than an order of magnitude: a reduction of kernel cycles of \num{16.55(23)}$\times$ and \num{14.72(30)}$\times$ speed-up in optical photon simulation wall time including overheads, at the reorder interval $N=4$, relative to the same one thread per photon megakernel in a \SI{14.7}{\kilo\tonne} liquid argon time projection chamber. It resulted in bit identical output. Overheads included data movement from host to GPU, GPU initialization and copying back the hits to the host. The largest measured improvement is the recovery of active warp lanes from \num{2.64} to \num{20.38} of 32. Together with the improvement in predicated instruction issue and the occupancy driven increase in issue slot activity, it is consistent with the observed cycle reduction to within ${\sim}6\%$. Meanwhile branch efficiency changes negligibly, and we observe no evidence of a memory bandwidth limit. A control build with the reorder call compiled in but never executing separates the speed-up associated with the pipeline and launch configuration changes from that due to executed reordering: the former contributes $3.2\times$, and the latter contributes a further $5.1\times$.

We additionally demonstrated that when replacing ray tracing with analytical navigation in the bulk liquid argon volume and supplying a coherence hint that groups photons by the volume they currently occupy, it speeds up the simulation by a further ${\sim}13\%$. The hint keeps ray traced and analytically transported photons in separate warps, mitigating the branch divergence introduced by this hybrid method. Finally, we showed that the GPU energy consumption drops by ${\sim}90\%$ for a typical high energy workload by enabling SER.

This technique could accelerate other execution divergent, heavy tailed ray tracing workloads, such as mesh based Monte Carlo photon transport in biological tissue in medical physics, and potentially other applications of ray tracing as well. The hybrid result also suggests a route for analytic transport codes such as IceCube's: analytic stepping hosted in a ray generation shader, hint only reordering in place of the photon pool approach, and ray traversal engaged only near detector geometry.

\section*{Acknowledgements}
This work was supported by the Laboratory Directed Research and Development (LDRD) program of Brookhaven National Laboratory under project 26794.

\section*{Declaration of competing interest}
The author declares no competing interests.

\section*{Data availability}
Simphony is open source and available at \url{https://github.com/BNLNPPS/simphony}. The benchmark geometry, optical property tables, and analysis scripts are available from the author on reasonable request; the underlying optical simulation and its validation against Geant4 are described in another paper~\cite{companion}.

\end{document}